\documentclass{article}
\usepackage{ijcai22}

\usepackage{times}
\usepackage{soul}
\usepackage{url}
\usepackage[hidelinks]{hyperref}
\usepackage[utf8]{inputenc}
\usepackage[small]{caption}
\usepackage{graphicx}
\usepackage{natbib}
\title{Human-AI communication for human-human communication: \\Applying interpretable unsupervised anomaly detection to executive coaching}

\author{
Riku Arakawa\footnote{These authors contributed equally and are ordered alphabetically.}$^1$
\And
Hiromu Yakura$^{*2,3}$
\affiliations
$^1$Carnegie Mellon University, Pittsburgh, USA\\
$^2$University of Tsukuba, Tsukuba, Japan\\
$^3$National Institute of Advanced Industrial Science and Technology (AIST), Tsukuba, Japan
\emails
rarakawa@cs.cmu.edu,
hiromu.yakura@aist.go.jp
}

\newcommand{\figref}[1]{Figure~\ref{#1}}
\newcommand{\secref}[1]{Section~\ref{#1}}

\newcommand{\REsCUE}{REsCUE}
\newcommand{\INWARD}{INWARD}

\begin{document}

\maketitle

\begin{abstract}
In this paper, we discuss the potential of applying unsupervised anomaly detection in constructing AI-based interactive systems that deal with highly contextual situations, i.e., human-human communication, in collaboration with domain experts.
We reached this approach of utilizing unsupervised anomaly detection through our experience of developing a computational support tool for executive coaching, which taught us the importance of providing interpretable results so that expert coaches can take both the results and contexts into account.
The key idea behind this approach is to leave room for expert coaches to unleash their open-ended interpretations, rather than simplifying the nature of social interactions to well-defined problems that are tractable by conventional supervised algorithms.
In addition, we found that this approach can be extended to nurturing novice coaches; by prompting them to interpret the results from the system, it can provide the coaches with educational opportunities.
Although the applicability of this approach should be validated in other domains, we believe that the idea of leveraging unsupervised anomaly detection to construct AI-based interactive systems would shed light on another direction of human-AI communication.
\end{abstract}

\section{Introduction}

Social signal processing is one of the prominent application areas of machine learning \citep{DBLP:journals/taffco/VinciarelliPHPPDS12,DBLP:books/cu/p/RudovicNP17}.
Conventionally, supervised learning has been leveraged in combination with handcrafted features to obtain useful information from human conversation data \citep{DBLP:journals/tmm/Sanchez-CortesAMG12,DBLP:journals/tmm/BeyanCBM18}.
For example, \citet{DBLP:journals/tmm/Sanchez-CortesAMG12} designed various audiovisual features, such as the number of segments where the amount of one's body movement exceeded a certain threshold, to detect emergent leaders within a group discussion.
Rule-based methods have also been employed, especially to provide feedback based on social signal processing \citep{DBLP:conf/chi/TausczikP13,DBLP:conf/chi/DamianTBSLA15,DBLP:conf/iui/TanveerLH15,DBLP:conf/icmi/DamianBA16}.
In this case, both conversation \citep{DBLP:conf/chi/TausczikP13,DBLP:conf/icmi/DamianBA16} and public speaking \citep{DBLP:conf/chi/DamianTBSLA15,DBLP:conf/iui/TanveerLH15} have been dealt with.


However, such approaches based on some rules or supervised learning have a risk of oversimplifying our nature of social interaction, which is highly contextual \citep{RePEc:nbr:nberwo:0042}.
Indeed, most of us would agree that keeping one's arms crossed for a while in public speaking constitutes a negative behavior.
Thus, giving users feedback to refrain from doing it, in a similar manner as \citet{DBLP:conf/chi/DamianTBSLA15} did, would be beneficial.
However, in everyday conversation, we cannot conclude that crossed arms immediately represent a defensive attitude \citep{Navarro2008}, because it might be a sign of one's perseverance in finding an answer to a given question \citep{10.1002/ejsp.444}.

A more extreme case is given by executive coaching, in which coaches are asked to observe the nonverbal behavior of coachees during one-on-one sessions \citep{isbn:9780761939764}.
Here, coaches are required to infer a coachee's internal status from the nonverbal behavior of the coachee to complement the verbal communication \citep{cox2009complete}.
In particular, the importance of observing such nonverbal behavior is emphasized in terms of reading the nuance of what the coachee said (and not said) \citep{drake2009narrative}.
In this case, na\"ive rules or simplified classifications used with supervised learning would not be sufficient to construct computational support or might even misguide coaches in capturing the complex dynamics of humans.

Given this background, we have developed a practical approach for supporting executive coaching by exploiting unsupervised anomaly detection to provide coaches with interpretability.
Based on our experiences, in this paper, we discuss the importance of securing interpretability in implementing collaboration between AI and experts within such highly contextual situations.
In addition, we present the educational effect of designing a computational support tool with the approach based on our observation about how this type of human-AI communication can affect human-human communication.

\section{Why unsupervised anomaly detection?}
\label{sec:why-unsupervised}

As mentioned above, human conversation, particularly in executive coaching, is highly context dependent \citep{watzlawick2011pragmatics,cox2009complete}.
This point broke our initial attempt to apply supervised learning algorithms in developing computational support.
Specifically, we first deployed existing supervised-algorithm-based methods for recognizing emotions \citep{DBLP:conf/icmi/DhallGGJHG17} and detecting important utterances \citep{DBLP:conf/icmi/NiheiNT17} in conversation during executive coaching sessions.
However, we found that expert coaches begin to ignore the outputs from a computer once the outputs contradict their observation or intuition.
As per their comments, the coaches found it difficult to rely on outputs based on simplified categories that are indifferent to the subtle changes in the nonverbal behavior of the coachees, leading them to disregard the computer.

With this finding, we developed \textit{\REsCUE{}} \citep{DBLP:conf/chi/ArakawaY19}, an alternative approach that harnesses multimodal anomaly detection.
\REsCUE{} is designed to detect subtle but meaningful changes in a coachee's nonverbal behavior that seem to be outliers from their behavior trends.
It then presents the changes to a coach as cues to infer the coachee's internal status.
Then, the coach can read what exists behind the detected cues by considering the context of their conversation.
This approach frees us from implicit assumptions about human nonverbal behavior that have been introduced by heuristic rules or training data in conventional methods.

\begin{figure}[t]
    \centering
    \includegraphics[width=\linewidth]{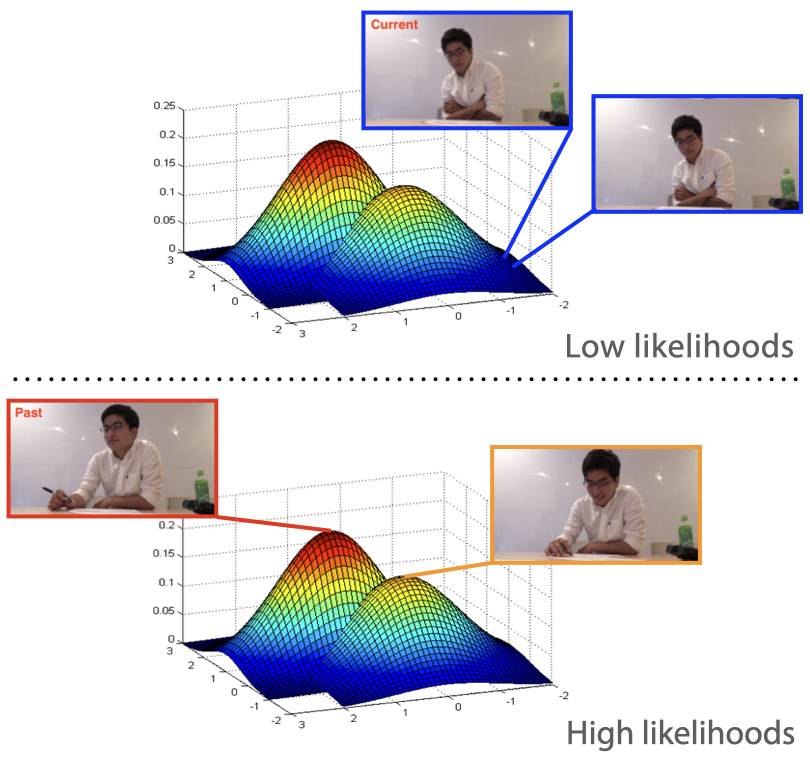}
    \caption{\textit{\REsCUE{}} \protect\citep{DBLP:conf/chi/ArakawaY19} can determine changes in nonverbal behavior along with past representative states using GMM.}
    \label{fig:gmm}
\end{figure}

\begin{figure}[t]
    \centering
    \includegraphics[width=0.9\linewidth]{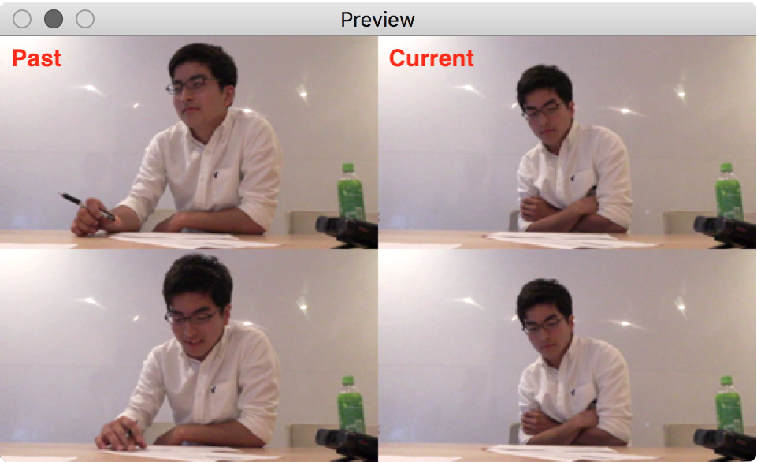}
    \caption{The detected nonverbal cues are visualized in an interpretable manner by being supplied with past representative states (taken from Figure 6 of \protect\citep{DBLP:conf/chi/ArakawaY19}).}
    \label{fig:visualize}
\end{figure}

Furthermore, the anomaly detection algorithm of \REsCUE{}, which utilizes the Gaussian mixture model (GMM), enables an interpretable presentation of the nonverbal cues.
As illustrated in \figref{fig:gmm}, \REsCUE{} can determine the moments that the coachee exhibited the most outlying behavior, which has the minimum likelihood according to the GMM.
At the same time, it can determine the moments that the coachee exhibited representative states among their past behavior analogously by looking for those having the maximum likelihood.
Then, just by arranging them as in \figref{fig:visualize}, \REsCUE{} can visualize the cue so that the coach can intuitively interpret its implications during one-on-one sessions.

The design of \REsCUE{} is principled in the separation of observation and interpretation.
From our initial attempt, we learned that expert coaches in fact do not have difficulty in interpreting the implications of nonverbal cues they observed.
Rather, they are afraid of missing out on some cues while mentally demanded to manage sessions by, for example, finding an appropriate question that can provide a chance for coachees to have an effective reflection.
Thus, we designed \REsCUE{} to devote itself to supporting only the observation part and delegate the interpretation part to coaches via visualizing the observed cues in an interpretable manner, instead of directly inferring the meaning of the cues based on rules or simplified classifications. 
The effectiveness of this design was experimentally confirmed, and now, \REsCUE{} is practically deployed as a supporting tool for executive coaching.

\section{Lens of Parasuraman's framework of automation}

In the context of human-AI communication, the advantage of \REsCUE{} can be understood by using the framework of \citet{DBLP:journals/tsmc/ParasuramanSW00}, which provides an objective basis for automation design.
They presented that automation can be applied to four classes of functions: information acquisition, information analysis, decision \& action selection, and action implementation.
Then, they defined a 10-point scale to evaluate the automation level for each function, from level 1 (``the computer offers no assistance: human must take all decisions and actions'') to level 10 (``the computer decides everything, acts autonomously, ignoring the human'').
This framework allows us to assess the suitability of automation design by determining whether the automation level of each function is appropriately situated within the trade-off between human performance, automation reliability, and cost of consequences.

\citet{DBLP:journals/tsmc/ParasuramanSW00} takes an example of the ground proximity warning system (GPWS) that alerts pilots when their aircraft is flying into the ground.
They classified it as level 4 (``suggests one alternative'') in decision \& action selection because it does not execute any countermeasures but makes suggestions for pilots with an audio message saying ``pull up!''
However, when such an alert is triggered, the performance of pilots is often degraded but its cost of consequences can be too high, i.e., resulting in a crash.
Therefore, GPWS is now being replaced with an automatic ground collision avoidance, which exhibits level 7 (``executes automatically, then necessarily informs the human'') automation by taking control automatically if the pilot does not.

In this framework, \REsCUE{} shows its distinctive design approach; it retains level 10 for both information acquisition and information analysis while exhibiting level 1 for both decision \& action selection and action implementation.
Specifically, \REsCUE{} acquires the information about a coachee's nonverbal behavior and analyzes its outlierness, working independently from the coach.
However, it touches neither decision \& action selection nor action implementation; it leaves the coach to interpret the observed cues and make a necessary action (e.g., asking the coachee another question to figure out the reason behind it).

This is rationalized by considering the trade-off between human performance, automation reliability, and cost of consequences.
As mentioned in \secref{sec:why-unsupervised}, coaches' performance in observing coachees' nonverbal behavior is not consistent due to the mental demand of carrying out the sessions.
On the contrary, we found that the reliability of \REsCUE{} in detecting nonverbal cues is quite high because it does not introduce any heuristics or biases caused by training data.
Furthermore, its interpretable visualization (see \figref{fig:visualize}) allows the coach to ignore the cues when they are not implicational, which minimizes the negative cost of consequences.

A similar discussion can be applied to the interpretation part, i.e., high human performance with low automation reliability.
In addition, the cost of consequence is higher than the observation part because misinterpretation of the nonverbal cues can mislead the session.
To sum, the reason behind coaches' affirmative reception of our approach can be attributed to this automation design strategy.
This suggests that the clear distinction between the observation and interpretation that is enabled by unsupervised anomaly detection can support human-AI communication in highly contextual situations.

\section{Asking humans to explain}

Following the deployment of \REsCUE{}, we found that this approach can be applied to supporting the education of novice coaches.
This is because the ability to infer one's internal status from such nonverbal cues, which is often associated with emotional intelligence \citep{doi:10.2190/DUGG-P24E-52WK-6CDG}, plays an important role in successful executive coaching \citep{Wasylyshyn2012}.
However, intensive training is often required in cultivating such an ability \citep{PatMcEnrue2009}.
We thus presumed that \REsCUE{} can help the transfer of the ability from expert coaches to novice coaches, leading us to develop a computational supporting tool: \textit{\INWARD{}} \citep{DBLP:conf/chi/ArakawaY20}.

\begin{figure}[t]
    \centering
    \vspace{10pt}
    \includegraphics[width=\linewidth]{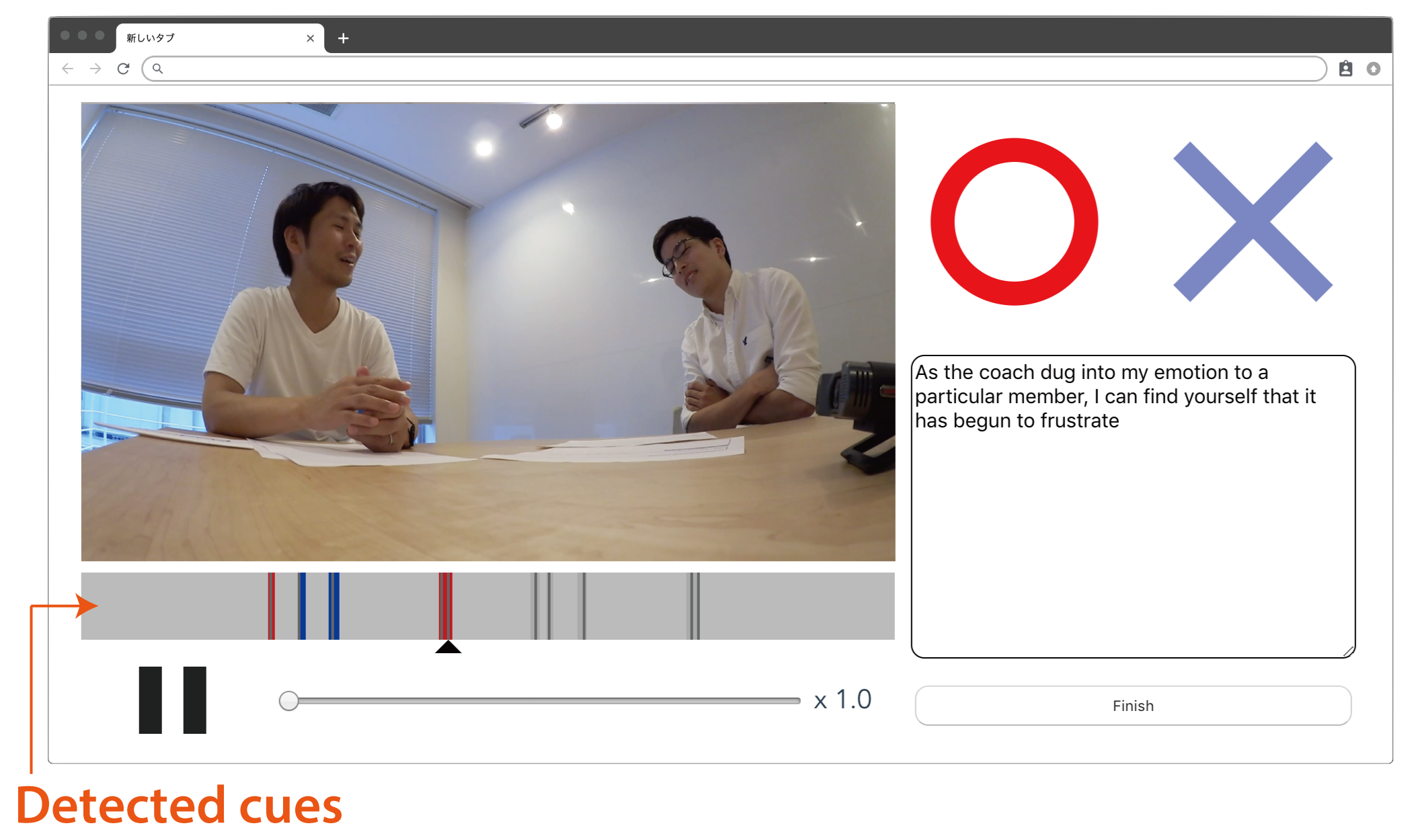}
    \caption{\textit{\INWARD{}} \protect\citep{DBLP:conf/chi/ArakawaY20} provides an interface for reflecting on coaching sessions with detected nonverbal cues.}
    \vspace{15pt}
    \label{fig:inward-video}
\end{figure}

\begin{figure}[t]
    \centering
    \includegraphics[width=\linewidth]{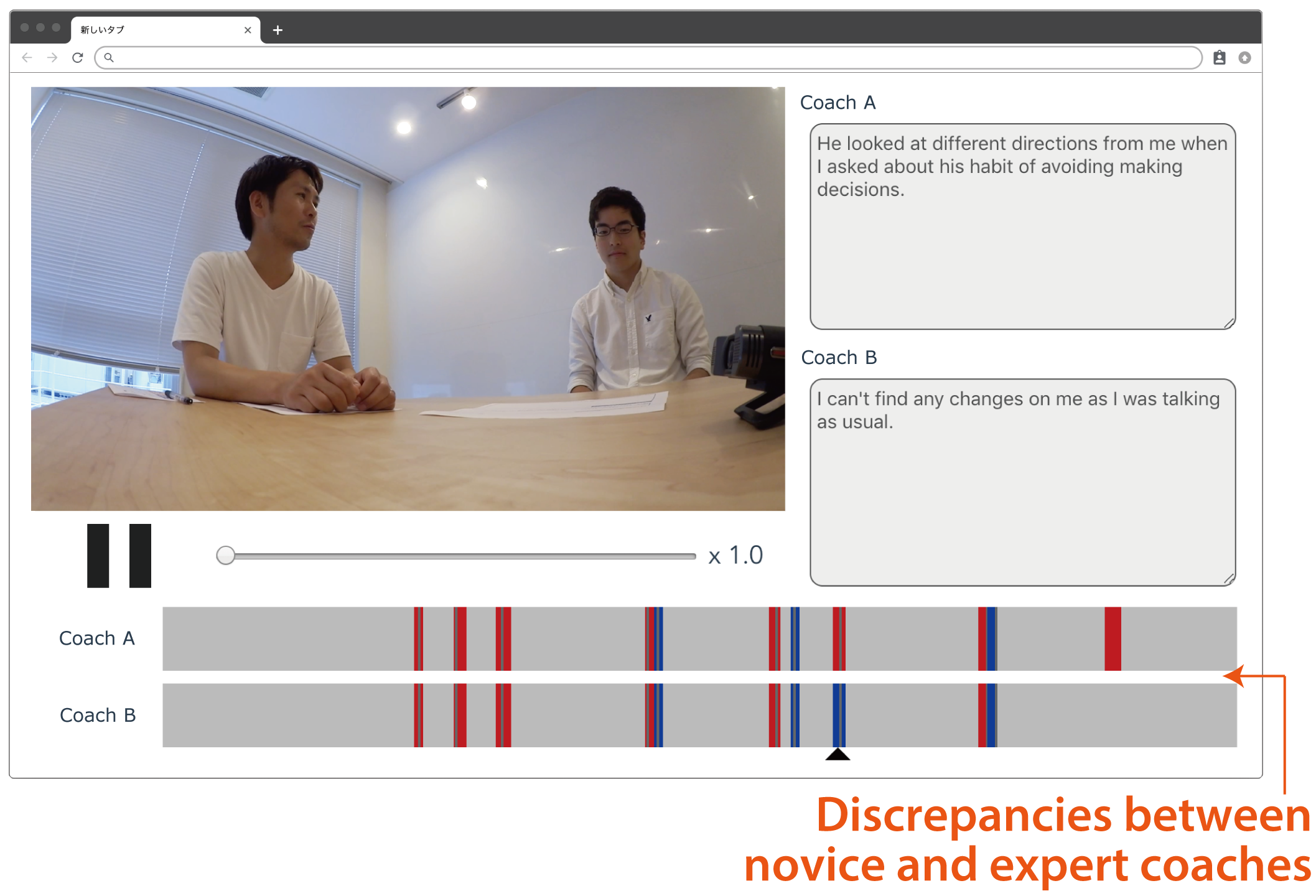}
    \caption{\textit{\INWARD{}} \protect\citep{DBLP:conf/chi/ArakawaY20} then visualizes the discrepancies in the interpretations of the coaches to encourage their meta reflection.}
    \label{fig:inward-meta}
\end{figure}

As shown in \figref{fig:inward-video}, this tool provides an interface displaying the recording of a coaching session as well as nonverbal cues detected by \REsCUE{} \citep{DBLP:conf/chi/ArakawaY19}.
It assumes that a novice coach who had the session and an expert coach who supervises the novice coach use the interface separately to reflect on the session.
In this video reflection phase, the coaches are requested to annotate each detected cue whether or not it provides insights about the coachee's internal status and to explain the reason for such an annotation.
After that, this tool invites both the coaches to a meta reflection phase, in which the discrepancies between their annotations are displayed, as depicted in \figref{fig:inward-meta}, so that they can discuss their interpretations.
Through these phases, the novice coach can learn how the expert coach reads information from subtle nonverbal cues.

Here, we cannot deny the possibility that the novice coach can learn the way of interpreting such nonverbal cues equivalently without the computational support of \INWARD{}; for example, they can watch the recording along with the expert coach while being explained about the interpretations of the expert coach.
However, the user study of \INWARD{} revealed that its detection results serve as a neutral ground for their discussion.
Specifically, during the video reflection phase, the novice coach can construct their own hypothesis regarding the tendency in the coachee's nonverbal behavior through interpreting and annotating each detected cue.
Subsequently, the novice coach can argue with the expert coach without unthinkingly accepting the interpretations of the expert coach on account of authority bias.

In addition, we found that the design of \REsCUE{}, which separates observation and interpretation, fosters the effectiveness of \INWARD{}.
As \REsCUE{} is based on unsupervised anomaly detection, it can detect apparent behavioral changes, such as when the coachee takes their personal organizer out of a briefcase.
Most of them would not provide meaningful information about the coachee's internal status, but occasionally, expert coaches draw insights with regard to why the coachee did it at the moment.
Such open-ended interpretations would not be available unless the tool completely delegates the interpretation part to humans while providing interpretable results.

Through the development of \INWARD{}, we learned another advantage of making AIs interpretable, i.e., they can offer educational opportunities for humans by actively asking them to interpret and explain along with supplying interpretable results.
Needless to say, the transparency of the AIs is crucial not to make humans feel quizzed by computers.
Besides, we found that unsupervised algorithms would have an affinity to highly contextual situations, such as executive coaching, as they can induce open-ended interpretations from humans.
We believe that applying this approach to other domains sheds light on a new direction of AI techniques in interpersonal communication.

\section{On-going work}

\begin{figure}[t]
    \centering
    \includegraphics[width=\linewidth]{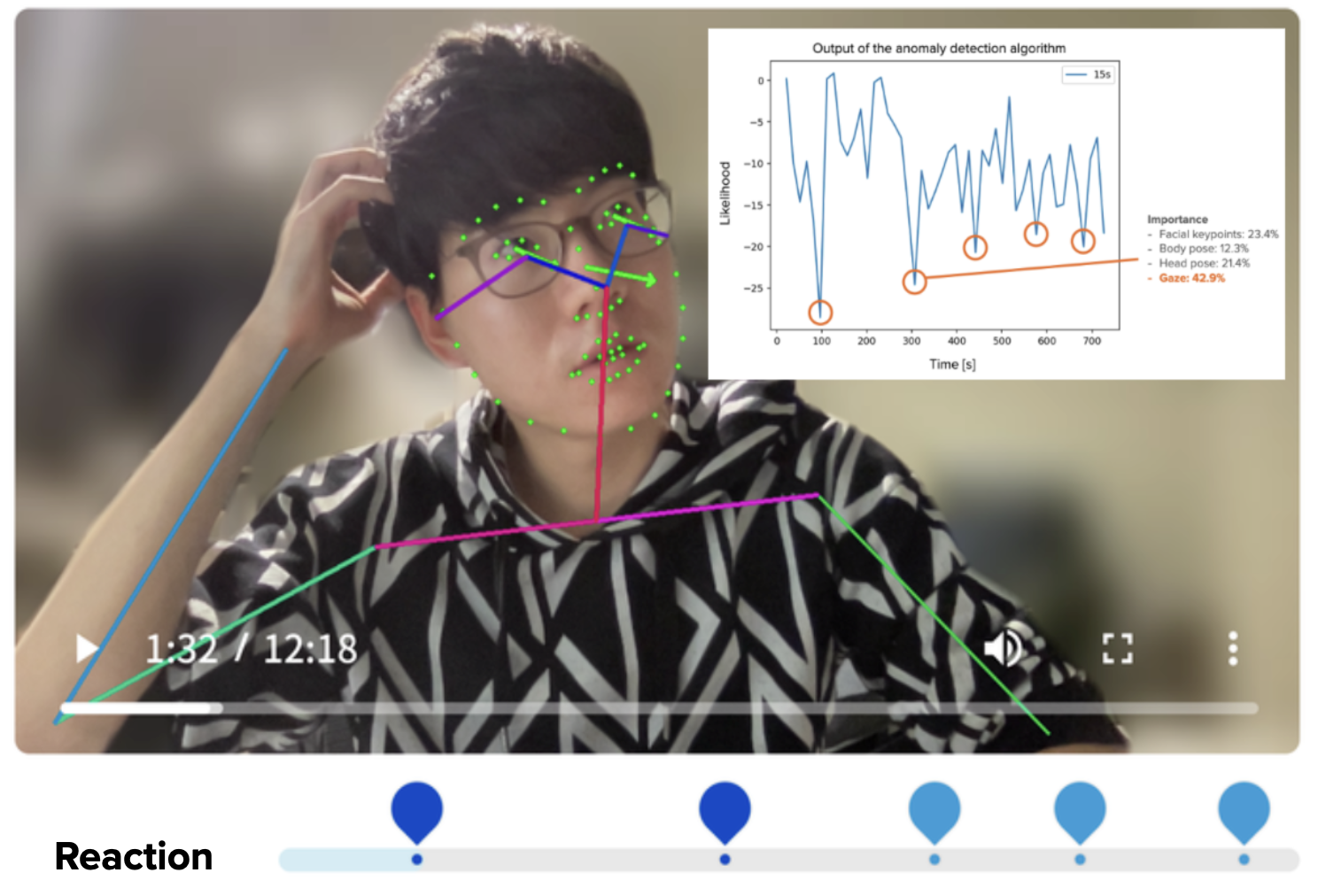}
    \caption{Prototype being developed for the ongoing work: supporting assessors in human assessment.} 
    \label{fig:new-interface}
\end{figure}

To this end, we are currently working on applying this approach based on unsupervised anomaly detection in another domain of human communication, that is, human assessment.
Human assessment is a process in human resource development aimed to evaluate and make decisions on candidates regarding their suitability for certain types of employment \citep{Sartori2020Psychological}.
Typically, professional assessors have an interview session with candidates through which the candidates' skills as a manager in their organizations are examined \citep{Roberts2014TheValidity}.
The assessors are required to monitor and evaluate the candidates' behavior during the interview.
This process is similar to executive coaching in terms of analyzing one's behavior during a one-on-one conversation, involving highly contextual situations.
At the same time, it is different in the sense that the goal of human assessment is to make a judgment on candidates' suitability for jobs while executive coaching focuses on nurturing coachees by having them deeply reflect on their internal states.
As a result, the ongoing work possesses a unique challenge with respect to how to support professionals' decision-making with AIs.
Our hypothesis is that our approach of unsupervised anomaly detection can detect scenes that can be served as a basis for the assessors to evaluate candidates, making the process more effective and efficient.
If this is supported through empirical studies, the boundary of this approach's capability regarding what types of scenes the anomaly detection algorithm can capture within human communication will be clearer.
To conduct such studies, we are developing a prototype interface as shown in \figref{fig:new-interface}, while exploring useful features to be used for anomaly detection in a collaboration with professional assessors.

\section{Conclusions}

In this paper, we have discussed the importance of deploying interpretable AIs in highly contextual situations, such as human conversation, by introducing our cases in executive coaching.
In these situations, we argue that it would be worth thinking about delegating the interpretation part to humans, instead of persuading them by presenting possible explanations, while fully supporting their interpretation by providing interpretable outputs.
This approach also has the potential to provide users with educational opportunities by intentionally requesting them to interpret the outputs, as we discussed by taking an example of \INWARD{} \citep{DBLP:conf/chi/ArakawaY20}.

In this sense, we envision that unsupervised anomaly detection will be utilized more widely in the future as an alternative approach in constructing human-AI communication for effective collaboration.
As discussed above, human nature is often too complex to be simplified into well-defined problems tractable by regular supervised learning algorithms.
Then, applying the existing explainable AI techniques designed for supervised algorithms to such a situation might not be a remedy.
Instead, when we can assume the involvement of experts as the counterpart of AIs, we suggest delegating the interpretation part to them, namely, trusting them, to induce their open-ended interpretations.

\section*{Acknowledgments}

This work was supported in part by JST ACT-X Grant Number JPMJAX200R and JSPS KAKENHI Grant Number JP21J20353.

\bibliographystyle{named}
\bibliography{ijcai22}

\end{document}